\documentclass[preprint,prd,tightenlines,nofootinbib]{revtex4}

\pdfoutput=1
\usepackage{amsmath,bm}
\usepackage{graphicx}
\usepackage{hyperref} 

\usepackage{color}
\newcommand{\LA}{\mathrm{A}}
\newcommand{\LB}{\mathrm{B}}

\newcommand{\LS}{\mathrm{S}}
\newcommand{\LT}{\mathrm{T}}
\newcommand{\LW}{\mathrm{W}}
\newcommand{\LZ}{\mathrm{Z}}
\newcommand{\La}{\mathrm{a}}
\newcommand{\Lb}{\mathrm{b}}

\newcommand{\Lg}{\mathrm{g}}
\newcommand{\Lp}{\mathrm{p}}

\newcommand{\Lt}{\mathrm{t}}
\newcommand{\Lu}{\mathrm{u}}

\newcommand{\GeV}{\ \mathrm{GeV}}

\newcommand{\tildenot}{{\raise.17ex\hbox{$\scriptstyle\mathtt{\sim}$}}} % "~"

\begin{document}

\title{Finding physics signals with event deconstruction}

\author{Davison E. Soper}
\affiliation{
Institute of Theoretical Science\\
University of Oregon\\
Eugene, OR  97403-5203, USA\\
}

\author{Michael Spannowsky}
\affiliation{
Institute for Particle Physics Phenomenology\\
Department of Physics\\
Durham University\\
Durham CH1 3LE,
United Kingdom\\
}

\begin{abstract}
We develop a matrix element based reconstruction method called event deconstruction. The method uses information from the hard matrix element and a parton shower to assign probabilities to whether a final state was initiated by a signal or background process. We apply this method in the signal process of a $Z'$ decaying to boosted top quarks in an all hadronic final state and discuss envisioned improvements of the method. We find that event deconstruction can considerably improve on existing reconstruction techniques.
\end{abstract}

%\keywords{perturbative QCD, parton shower}
\preprint{IPPP/14/04}
\preprint{DCPT/14/08}

\maketitle
%-------------------------------------------------------------------

\section{Introduction}

After the recent discovery of a Standard Model-like Higgs boson \cite{Aad:2012tfa,Chatrchyan:2012ufa}, the two main goals of the upcoming LHC runs are to measure the properties of this resonance and to look for signals for physics beyond the Standard Model. In the previous run, new physics signals remained elusive. This may indicate that the mass scale of new physics is very high. Alternatively, it may result from limitations on our ability to disentangle the signals from large Standard Model backgrounds. Therefore, developing new tools designed to improve on the sensitivity in analyzing large data samples are of utmost importance for the success of the LHC experiments. To help with this task, we propose a method, event deconstruction, that aims to assign a likelihood ratio to each event that distinguishes whether the event was likely to have been created by a certain signal process or likely to have been created by a background process. The likelihood ratio is calculated from the matrix element for the hard interaction at the heart of the process combined with a parton shower model for softer interactions.

Event deconstruction is an extension of the shower deconstruction method \cite{SSHiggs,SStop} that we have introduced earlier. In shower deconstruction, we analyze the contents of a jet defined with a large jet radius $R$ (using, for example, the Cambridge-Aachen algorithm \cite{Cambridge, Aachen}). The object is to determine a likelihood ratio $\chi$ that tells us whether this fat jet contains something interesting that we are looking for, like the decay products of a top quark or a Higgs boson, or whether it is more likely to be an ordinary QCD jet. In event deconstruction, we wish to construct a likelihood ratio $\chi$ that tells us whether a whole event is likely to have come from an interesting signal like a $\LZ'$ boson that decays to $\Lt + \bar \Lt$, or whether it came from a prosaic background process, ordinary QCD jets or QCD production of $\Lt + \bar \Lt$. We analyze two or more fat jets or, ideally, a whole event. We include in the analysis the probabilities for the signal or background hard processes that create the hard jets. These probabilities are calculated from the squared matrix elements times the proper parton distribution functions. 

The event deconstruction method bears some resemblance to the well known and widely used matrix element method \cite{MEM1, MEM2, MEM3, MEM4, MEM5, MEM6, MEM7, MEM8, MEM9, MEM10}.
 
We organize this work as follows: First we describe the way we implement event deconstruction in Sec.~\ref{sec:ED}. In Sec.~\ref{sec:MEM}, we compare this method to the matrix element method. Then in Sec.~\ref{sec:ttbar} we test event deconstruction in a specific example. Finally, in Sec.~\ref{sec:outlook}, we present an outlook emphasizing avenues for improvement in the method.

\section{Event deconstruction}
\label{sec:ED}

Suppose that we have in mind a specific signal process. We wish to determine from data whether this signal process occurs in nature. Other processes in the Standard Model that may produce events that resemble signal events constitute possible backgrounds. For each experimental event, we can entertain the signal hypothesis that the event arose from the signal process and we can consider also the background hypothesis that the event arose from one of the background processes. For instance, the signal hypothesis might be that the hard process in the event is $\Lp + \Lp \to \LZ' \to \Lt + \bar \Lt$ with hadronic decay of the top quarks. Other signal hypotheses could include Higgs bosons or SUSY particles. If the signal process is $\Lp + \Lp \to \LZ' \to \Lt + \bar \Lt$, then the dominating background processes include QCD two jet production with light quarks and gluons and also QCD $\Lt + \bar \Lt$ production. 

In this section, we consider a signal that leads to final state hadronic jets. One can consider final state leptons also, but then we would need a more elaborate notation.\footnote{If neutrinos or other invisible particles are produced, then some substantial revisions of the method are needed. We leave the investigation of this issue to future work.}

\subsection{The likelihood ratio $\chi$}
\label{sec:chi}

For each event, we group some subset of the hadrons in the event into jets according to, say, the Cambridge-Aachen jet algorithm \cite{Cambridge, Aachen}. We want a detailed picture of the event, so we use ``microjets'' with as small a value of the jet radius parameter $R$ as seems practical from an experimental point of view. Microjets with very small transverse momenta carry little useful information, so we eliminate those whose transverse momenta are judged to be too small. Thus, we work with the fine-grained information contained in the list $\{k\}_N = \{k_1,\dots,k_N\}$ of momenta of many microjets. With fine grained information, we may hope to do a good job of distinguishing signal from background. Additionally, since $R$ is small one can at least hope that the correspondence between microjet momenta and parton momenta is reasonably close.

To distinguish signal from background, we construct an approximate likelihood ratio
\begin{equation}
\label{eq:chi}
\chi = \frac{{\cal P}(\{k\}_N,{\rm signal})}{{\cal P}(\{k\}_N,{\rm background})}
\;,
\end{equation}
where ${\cal P}(\{k\}_N,{\rm signal})$ is the probability density to produce the observed microjet configuration according to the signal hypothesis and ${\cal P}(\{k\}_N,{\rm background})$ is the probability density to produce the microjet configuration according to the background hypothesis. Then we can test the signal hypothesis in a sample of experimental events by asking whether the number of events with large $\chi$ is greater than would be expected by chance if there were really no signal.

\subsection{Model hypotheses and event histories}
\label{sec:histories}

%---------------FIGURE------------------
\begin{figure}
\centerline{\includegraphics[width=11.0cm]{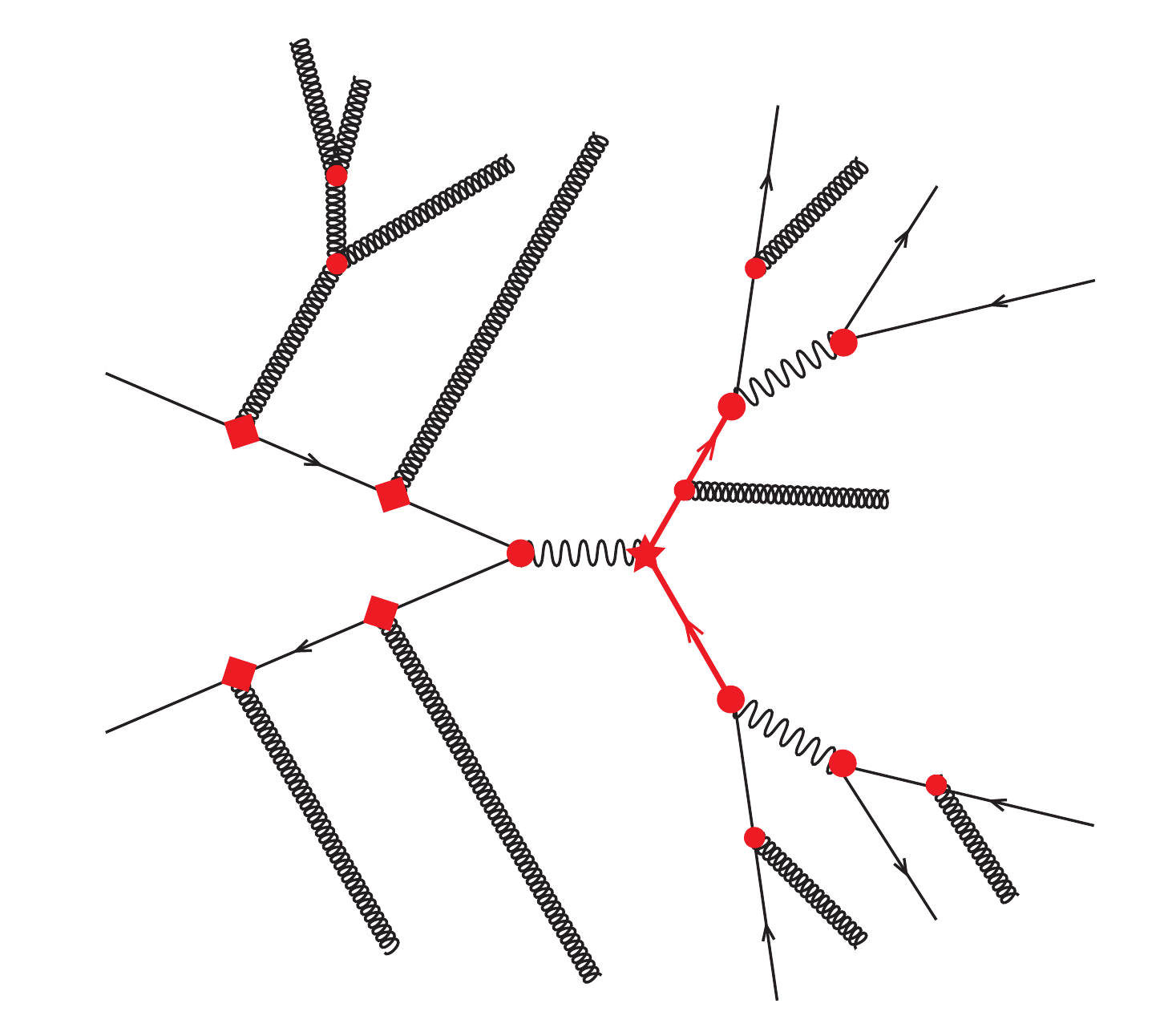}}
\caption{
A shower history for a $ \LZ' \to \Lt + \bar \Lt$ signal event. The top quarks decay hadronically. Initial state radiation is included as uncorrelated independent emissions. The vertices in the diagram represent QCD splittings in a shower approximation or decays with Breit-Wigner factors. The propagators of final state quarks and gluons represent Sudakov factors.
}
\label{fig:Shistory}
\end{figure}
%-------------END FIGURE----------------

In order to approximately calculate $\chi$, we need to calculate ${\cal P}(\{k\}_N,{\rm model})$, where ``model'' is the signal or background process. To see how to do this, let us suppose that we are interested in the signal process and that the signal is $\Lp + \Lp \to \LZ' \to \Lt + \bar \Lt$. Consider Fig.~\ref{fig:Shistory}, which depicts one event history according to the signal hypothesis. That is, Fig.~\ref{fig:Shistory} shows one way that a parton shower Monte Carlo event generator might have generated the event using the signal process. We identify the microjets with the partons at the end of the branchings in the diagram. There is an $s$-channel $\LZ'$ boson that decays to $\Lt + \bar \Lt$ and there are further decays of the top quarks and the $\LW$ bosons that they produce. Furthermore, the strongly interacting partons emit gluons. Additionally, there is initial state radiation from the incoming partons and there are QCD splittings of radiated gluons. With event deconstruction, we aim to approximate ${\cal P}(\{k\}_N,{\rm model})$ using the convolution of parton distribution functions and the squared matrix element for the hard process $q + \bar q \to \LZ' \to \Lt + \bar \Lt$ at the center of the diagram. Then we use a parton shower approximation for the rest of the diagram.

The use of a parton shower approximation is helpful for two reasons. First, it is difficult to calculate squared matrix elements at a high order of perturbation theory, such as that depicted in Fig.~\ref{fig:Shistory}. Second, some of the microjets may be nearly collinear with each other and some may be quite soft. Then we need the Sudakov factors \cite{sudakov:1956,Collins:1989bt} that effectively sum logarithms of the small angles or the ratios of a soft energy to a hard energy. These Sudakov factors are a normal part of a parton shower \cite{Herwig1, Herwig2, Pythia5, Pythia8, Sherpa, NSshower, showerreview}.

Both the hard matrix element and parton shower approximation may include an approximation for the decay of an unstable particle of flavor $f$, for instance a top quark or gauge boson. For this, we use a standard Breit-Wigner function with a resonance width $\Gamma(f)$. However, we are treating the momenta the partons at the end of the branchings in the diagram as if they were the same as the measured momenta of microjets. Really, the imperfections of this approximation together with experimental measurement errors should have the effect of convoluting the functions that we use with a smearing function, usually called the transfer function \cite{MEM1,MEM2}. If the functions that we use to calculate ${\cal P}(\{k\}_N,{\rm model})$ are slowly varying, this smearing should not matter. However, the Breit-Wigner functions can be quite sharply peaked. After smearing, the peak should be broader. We approximate the smeared Breit-Wigner functions by functions of the same form, but with a larger width $\Gamma'(f)$. The choice of $\Gamma'(f)$ is based on the experimentally resolvable width of the resonance.

When we use a parton shower approximation, Fig.~\ref{fig:Shistory} does not represent precisely a squared Feynman diagram. Rather, it represents an event history: the steps by which a parton shower event generator could have generated the given configuration of microjets. This includes Sudakov factors and quantum interference effects in a dipole shower.\footnote{Interference diagrams could also be included in the hard scattering for some processes.} There are many shower histories that can lead to the same configuration of microjets. Thus we should sum over event histories $h$:
\begin{equation}
\label{eq:simpleSD}
{\cal P}(\{k\}_N,{\rm model})
= \sum_h {\cal P}(\{k\}_N,{\rm model},h)
\end{equation}

Eq.~(\ref{eq:simpleSD}) expresses a conceptually simple ansatz for what event deconstruction should do. A parton shower event generator without hadronization and with a cutoff on how far shower evolution should go will generate configurations of partons that we can identify with microjets. The event generator uses probabilities based on parton distribution functions, a hard matrix element, parton splitting functions, and decay functions for unstable particles. It generates event histories $h$ at random according to these probabilities. Now, given the microjet configuration $\{k\}_N$, we use (an approximation to) the functions that are built into the event generator to calculate the probability that the given $\{k\}_N$ will be produced.

\subsection{Simplifying approximations}
\label{sec:approximations}

Eq.~(\ref{eq:simpleSD}) is conceptually simple, but computationally difficult. The basic problem is that if the number of microjets is large, the number of shower histories is very large. There are many ways to attack this problem of combinatorial explosion by making some suitable approximations. In this paper, we explore one method that is especially simple. In many cases of interest, the partons that are important for distinguishing signal from background are to be found in fairly small angular regions in the detector. Each of these angular regions contains the daughter partons from a high $p_\LT$ parton produced in the hard interaction. We can isolate the important regions by first analyzing the event for high $P_\LT$ jets, using a large value of the jet radius parameter $R$ of, say, the Cambridge-Aachen algorithm. In the case that the signal process is $\Lp + \Lp \to \LZ' \to \Lt + \bar \Lt$, we are looking for the possible daughter partons from the decay of the two top quarks. Thus we look for two ``fat'' jets each with a certain minimum $P_\LT$. In general, we would look for $J$ fat jets with labels $n = 1,\dots,J$.  

Having found the fat jets, we group the constituents of each fat jet into microjets using a jet algorithm with a small $R$. For fat jet $n$, this gives us a set of microjets with momenta $\{k^{(n)}_1,\dots,k^{(n)}_{N_n}\} = \{k\}_{N_n}^{(n)}$. The complete set of microjets is the union of these subsets.

We now simplify the calculation compared to that contemplated in Eq.~(\ref{eq:simpleSD}). According to a standard event generator picture, some of the microjets that make up each of the fat jets can arise as QCD splitting products of partons emitted as initial state radiation, as depicted in Fig.~\ref{fig:Shistory}. The probabilities for these initial state splittings are correlated with one another and, indeed, with initial state radiation that does not wind up in any fat jet. Our simplification is to treat each parton radiated from the initial state as a gluon whose emission is independent of all of the other emissions. Additionally, we do not identify which of the two initial state partons emitted the gluon. We have simply a model probability density $\rho(p_\LT)$ for radiating gluons from the initial state \cite{SSHiggs}. 

We also approximate the evolution of each fat jet as being independent of the evolution of the other fat jets. Thus, for instance, we set to zero the probability that a parton in one fat jet emits a soft gluon that winds up in another fat jet. Additionally, the internal dynamics of jet evolution involves a color coherence effect in which the angular distribution of gluon emission from one parton depends on the angle to a certain other parton, the color connected partner of the emitting parton \cite{Marchesini:1983bm}. But if there is no color connected partner within the fat jet, then we use an approximation \cite{SSHiggs} that does not make use of information about the color connected partner. Thus we do not ask whether a color connected partner might be found in one of the other fat jets.

\subsection{Shower probabilities for the fat jets}
\label{sec:shower}

With these simplifications, each fat jet consists of some initial state radiation together with decay and splitting daughters of a hard parton of flavor $f_n$ and momentum $p_n$. The internal development of each of the fat jets is independent of the development of the others.

Now we use a parton shower model to calculate approximately the probability ${\cal P}(\{k\}_{N_n}^{(n)},h_n,f_n)$ that the microjet configuration $\{k\}_{N_n}^{(n)}$ within fat jet $n$ arose from initial state radiation together with the decay of a high $p_\LT$ parton of flavor $f_n$ according to a shower history $h_n$. The shower history $h_n$ is the part of the complete event history that applies to fat jet $n$. To calculate the probability ${\cal P}$ , we use the method specified in refs.~\cite{SStop,SSHiggs}, omitting the factor that approximated the probability for a hard interaction to produce the hard parton. With event deconstruction, we will instead calculate the complete probability for the hard interaction. 

For each shower history $h_n$ for fat jet $n$, we know the momentum $p_n(h_n)$ of the hard parton of flavor $f_n$ that initiates the shower: it is the sum of the momenta of the microjets into which the hard parton splits. Note that $p_n(h_n)$ is not the total momentum of fat jet $n$ because some of this momentum is carried by initial state radiation. The initiating hard parton is normally not on shell: $p_n(h_n)^2 \ne m(f_n)^2$. However, $|p_n(h_n)^2 - m(f_n)^2|$ is typically much smaller than $|p_n(h_n)_\LT^2|$

\subsection{The event probability}
\label{sec:event}

With this setup, we can calculate an approximate probability for producing the complete microjet configuration $\{k\}_N$ according to the model of interest:
\begin{equation}
\begin{split}
{\cal P}(\{k\}_N,{\rm model})
={}& \sum_{ f_1,\dots,f_J}\sum_{h_1,\dots,h_J}
H(p(h_1),\dots,p(h_J);f_1,\dots,f_J)
\\&\quad\times
{\cal P}(\{k\}_{N_1}^{(1)},h_1,f_1)\,
\times \cdots \times
{\cal P}(\{k\}_{N_J}^{(J)},h_J,f_J)
\;.
\end{split}
\end{equation}
We sum over the possible flavors $f_n$ of the hard partons in our model. For example, for a background model with two fat jets, we might have $\{f_1,f_2\} = \{\Lu,\bar\Lu\}$, $\{\Lg,\Lg\}$, {\it etc}. Then, we sum over shower histories $h_n$ for each fat jet. For each combination of shower histories, we know the hard parton momenta $p_n(h_n)$. We use these to calculate the hard scattering cross section $H$ as a squared matrix element convoluted with parton distributions. The hard scattering cross section $H$ is typically derived as a function of dot products of the hard parton momenta $p_n(h_n)$, assuming that the $p_n(h_n)$ are on shell.  However, this is an approximation to the complete Feynman diagrams. We simply use the off-shell $p_n(h_n)$ in $H$. One could project the $p_n(h_n)$ to on-shell versions of these momenta, but we find that using the off-shell versions provides a better approximation to $(\sum_n p_n(h_n))^2$, which is important in the case that $\sum_n p_n(h_n)$ is the total momentum of a heavy resonance.

There is one subtlety to be mentioned for the calculation of $H$. The total momentum of the hard partons, $Q = \sum_n p_n(h_n)$, equals the total momentum of the two incoming partons: $Q = p_\La + p_\Lb$. The momenta $p_n(h_n)$ would determine the hard scattering function $H$ in a very standard way if the transverse part of $Q$ were zero.  However, there is initial state radiation, both in the fat jets and outside of them, that carries transverse momentum. Thus generally $Q_\perp \ne 0$. The incoming partons have momenta
\begin{equation}
\begin{split}
k_\La ={}& x_\La P_\LA + \frac{|k_{\La \perp}^2|}{2 x_\La P_\LA \cdot P_\LB}\, P_\LB + k_{\La \perp}
\;,
\\
k_\Lb ={}& \frac{|k_{\Lb \perp}^2|}{2x_\Lb P_\LA \cdot P_\LB}\, P_\LA + x_\LB P_\LB +  k_{\Lb \perp}
\;,
\end{split}
\end{equation}
where $P_\LA$ and $P_\LB$ are the momenta of the incoming hadrons, $x_\La$ and $x_\Lb$ are the momentum fractions of the incoming partons, and $k_{\La \perp}$ and $k_{\Lb \perp}$ are their transverse momenta. We know that $k_{\La \perp} + k_{\Lb \perp} = Q_\perp$ but we have no way of knowing $k_{\La \perp}$ and $k_{\Lb \perp}$ individually. To fix $k_{\La \perp}$ and $k_{\Lb \perp}$, we use the symmetric ansatz \cite{CSframe}
\begin{equation}
k_{\La \perp} = k_{\Lb \perp} = Q_\perp/2
\;.
\end{equation}
This fixes the kinematics.

We are thus able to calculate ${\cal P}(\{k\}_N,{\rm model})$ both for model = signal and for model = background. The ratio of these quantities is the likelihood ratio $\chi$ in Eq.~(\ref{eq:chi}).

\section{The matrix element method}
\label{sec:MEM}

Event deconstruction as presented above is quite similar to the matrix element method \cite{MEM1, MEM2, MEM3, MEM4, MEM5, MEM6, MEM7, MEM8, MEM9, MEM10}. In this section we point out some of the similarities and differences. 

To use the matrix element method, one still calculates a likelihood ratio similar to Eq.~(\ref{eq:chi}). For, say, the signal probability in the numerator, one convolutes parton distribution functions with the square of the sum of tree level Feynman diagrams like that in Fig.~\ref{fig:Shistory}. However, the order of perturbation theory represented in Fig.~\ref{fig:Shistory} is too high for a practical evaluation of Feynman diagrams. Thus one must use lower order perturbation theory and fewer jets. This means that the jets should be defined with a larger value of the radius parameter $R$. 

On one hand, using a larger $R$ value is an advantage. With a small $R$ one can get close to the collinear and soft singularities of perturbation theory, so that the Sudakov factors built into a parton shower treatment are important. With a large $R$ it is an adequate approximation to use just tree level Feynman diagrams.

On the other hand, using a large $R$ value and just a few jets entails a loss of information that may result in a loss of ability to distinguish between signal events and background events.

With a large value of $R$, the correspondence between parton momenta $k_i$ and the observed jet momenta $K_i$ is weakened. For this reason, one usually convolutes the probabilities ${\cal P}(\{k\}_N,{\rm model})$ with transfer functions $T_i(k_i,K_i)$ \cite{MEM1,MEM2}. This convolution involves numerical integrations, which can be expensive in computer resources.

In contrast to the matrix element method, event deconstruction includes a parton shower approximation in the likelihood ratio of Eq.~(\ref{eq:chi}), thereby exploiting information contained in the radiation generated during the event evolution from the hard interaction scale down to a scale of order 10 GeV. This allows one to include more information in the analysis by using more reconstructed objects. Particularly for boosted final state jets, which contain subjets with small angular separations, using a shower approximation results in more reliable weights in the calculation of the likelihoods for the signal and background hypotheses. Further, the small microjet radius reduces the need for transfer functions, thus retaining an acceptable speed for the algorithm. Experimental uncertainties affect the matrix element weight strongly if decays of narrow resonances are considered. These uncertainties can be absorbed without reintroducing a transfer function by replacing the resonance's physical width with the a broader effective width in the Breit-Wigner propagator.

\section{An application}
\label{sec:ttbar}

In order to test how well event deconstruction works, we apply it to a simple example. We search for a heavy $\LZ'$ resonance that decays to $\Lt + \bar \Lt$ that, in turn, decay to hadrons. This is the process depicted in Fig.~\ref{fig:Shistory}. We generate samples of signal and background events using \textsc{Pythia 8} \cite{Pythia8}. For the signal sample, the mass of the $\LZ'$ is 1500 GeV. We use two backgrounds, ordinary QCD dijet production and the QCD process $\Lp\Lp \to \Lt \bar{\Lt}$. We remove the invisible particles from the fully hadronized final state and use the remaining particles with $|y|<5.0$ as input for the Cambridge-Aachen jet-finding algorithm \cite{Cambridge, Aachen} as implemented in \textsc{Fastjet} \cite{FastJet} with $R=1.5$. For the fat jets we require $|y_j|<2.5$. To accept an event we require at least two jets  with $p_{T,j}>400$ GeV each. We then analyze the part of the event consisting of the two fat jets with the highest $p_{T,j}$. After event selection cuts we find a cross section of $1.7$ nb for the dijet background and $2.2$ pb for the $t\bar{t}$ background. Because we leave the $\LZ'$-quark coupling unspecified, the signal cross section is not fixed. Then we study what the smallest signal cross section is that we can exclude in the channel with fully hadronic top decays. We calculate the matrix elements of the hard process in event deconstruction using \textsc{MadGraph} 5 \cite{MadGraph}.

Event deconstruction uses reconstructed objects as input, {\it i.e.} isolated leptons, isolated photons or narrow jets, the so-called microjets. However, in this application all microjets are narrow jets and we veto events with one or more isolated leptons or photons. To construct the microjets, we use the Cambridge-Aachen-algorithm with $R=0.2$ on the fat jets' constituents. We remove microjets from the analysis unless $p_{T}^{\rm{micro}}>10.0$ GeV. If more than 9 microjets are present in a fat jet, we remove those with the lowest $P_T$ values until nine microjets remain.

%---------------FIGURE------------------
\begin{figure}
\centerline{\includegraphics[width=12.0cm]{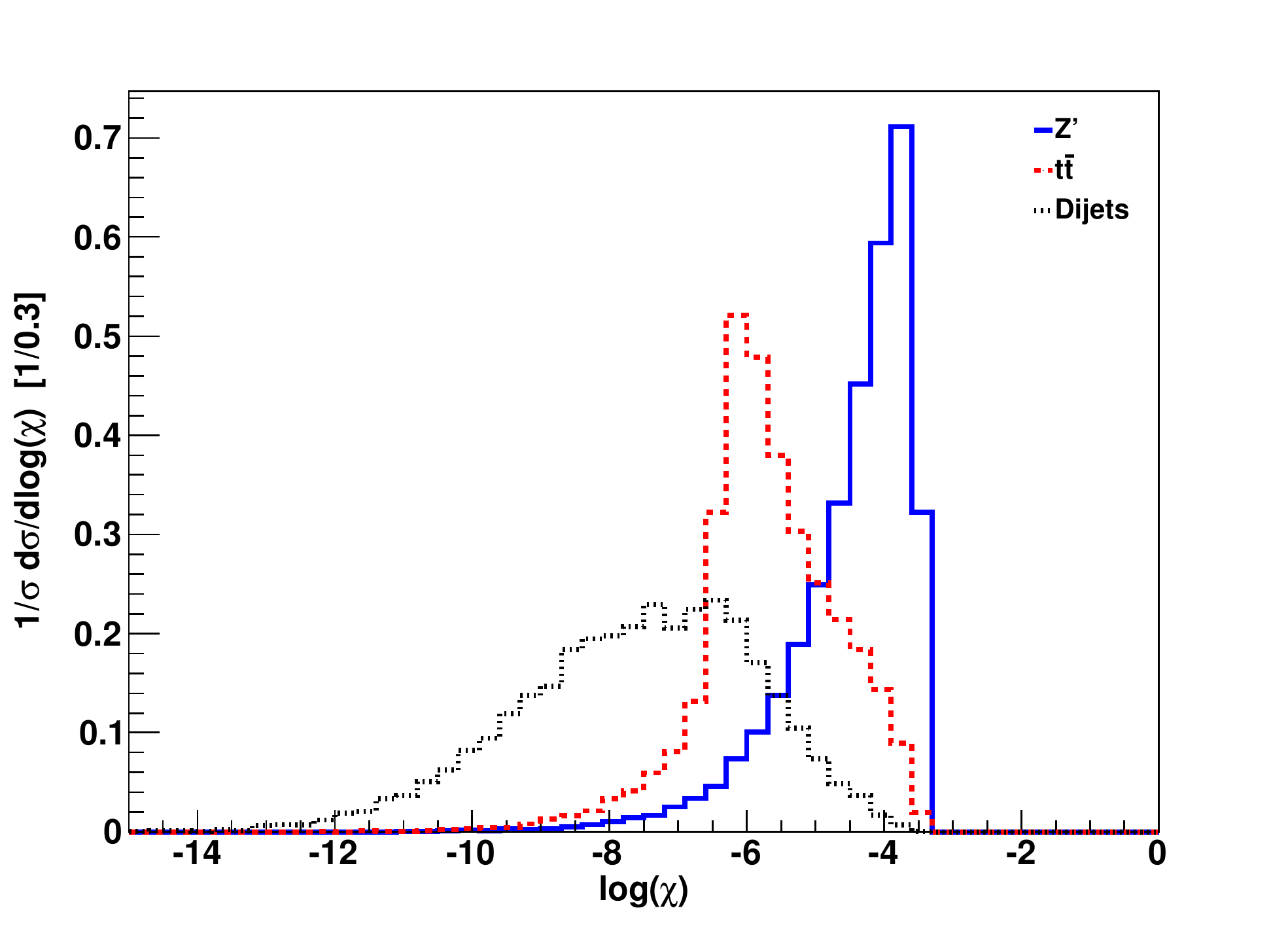}}
\caption{The $\log(\chi)$ distribution for the signal $Z'$ and the two background $t\bar{t}$ and dijets samples respectively. The events shown have passed the event selection cuts discussed in Sec.~\ref{sec:ttbar}. Events with $\chi=0$ are not shown.
}
\label{fig:chidists}
\end{figure}
%-------------END FIGURE----------------

For the signal model, we take $\Gamma(\LZ')_{\mathrm{model}} = 130 \GeV$. As we discussed in Sec.~\ref{sec:histories}, we have artificially increased the physical width assumed in generating signal events, $\Gamma(\LZ')_{\mathrm{phys}}=65 \GeV$, in order to account for inaccuracies in reconstructing the event. Similarly, we artificially set the top width to $\Gamma(\Lt) = 27 \GeV$ and the W width to $\Gamma(\LW) = 11 \GeV$. For the top and W-boson decays, we make the code more efficient by setting ${\cal P}(\{k\}_N,{\rm signal},h)$ to zero unless $|p^2 - M^2| < 2 M\Gamma$. 

We have also artificially decreased the $\LZ'$ mass in the event deconstruction to $M(\LZ')_{\mathrm{model}} = 1450 \GeV$ from the mass $M(\LZ')_{\mathrm{phys}} = 1500 \GeV$ in the events generated by \textsc{Pythia}. Although event deconstruction accounts for extra radiation entering the fat jets from initial state radiation and the underlying event, as described in \cite{SSHiggs}, the event deconstruction algorithm has no information about wide angle radiation that leaks out of the fat jets. This is a small effect that we have estimated by looking at the dijet mass distribution in signal events generated by \textsc{Pythia} with initial state radiation and the underlying event turned off. We find that the dijet mass distribution generated in this way peaks at $M(\LZ') = 1450 \GeV$ instead of $M(\LZ') = 1500 \GeV$. We have adjusted $M(\LZ')_{\mathrm{model}}$ to account for this effect. 

In the signal model, we assume that the $\LZ'$ couples equally to all flavors of quark with a vector-like coupling. We use simply 1 for the $\LZ'$-$q$ coupling. Using a different value simply shifts $\log \chi$ by a constant, the same constant for all events. Such a shift does not affect our analysis.

For each event in the signal sample and in each of the two background samples, we calculate $\chi$ according to Eq.~(\ref{eq:chi}). We can then plot the distribution of $\chi$ for each event sample. The resulting distributions are shown in Fig.~\ref{fig:chidists}. Some events do not show up in the plot because they have $\chi = 0$. In these events no microjet combination simultaneously satisfies all top and W mass window cuts. Excluding the $\chi = 0$ events, the distributions are normalized to $\int d\log\chi\ \rho(\chi) = 1$. We see that the $\chi$ distributions for the signal and the QCD dijet background have very different shapes, so that a simple cut on $\chi$ will do a good job of distinguishing the signal from this background. We see also that the shapes for the signal and the QCD $\Lt \bar \Lt$ background are not so different. Here the only distinguishing feature of the signal process is the presence of a $\LZ'$ resonance peak. Nevertheless, this distinguishing feature is sufficient to provide some discriminating power. Fortunately, the cross section for QCD $\Lt \bar \Lt$ production is quite small.

%---------------FIGURE------------------
\begin{figure}
\centerline{\includegraphics[width=12.0cm]{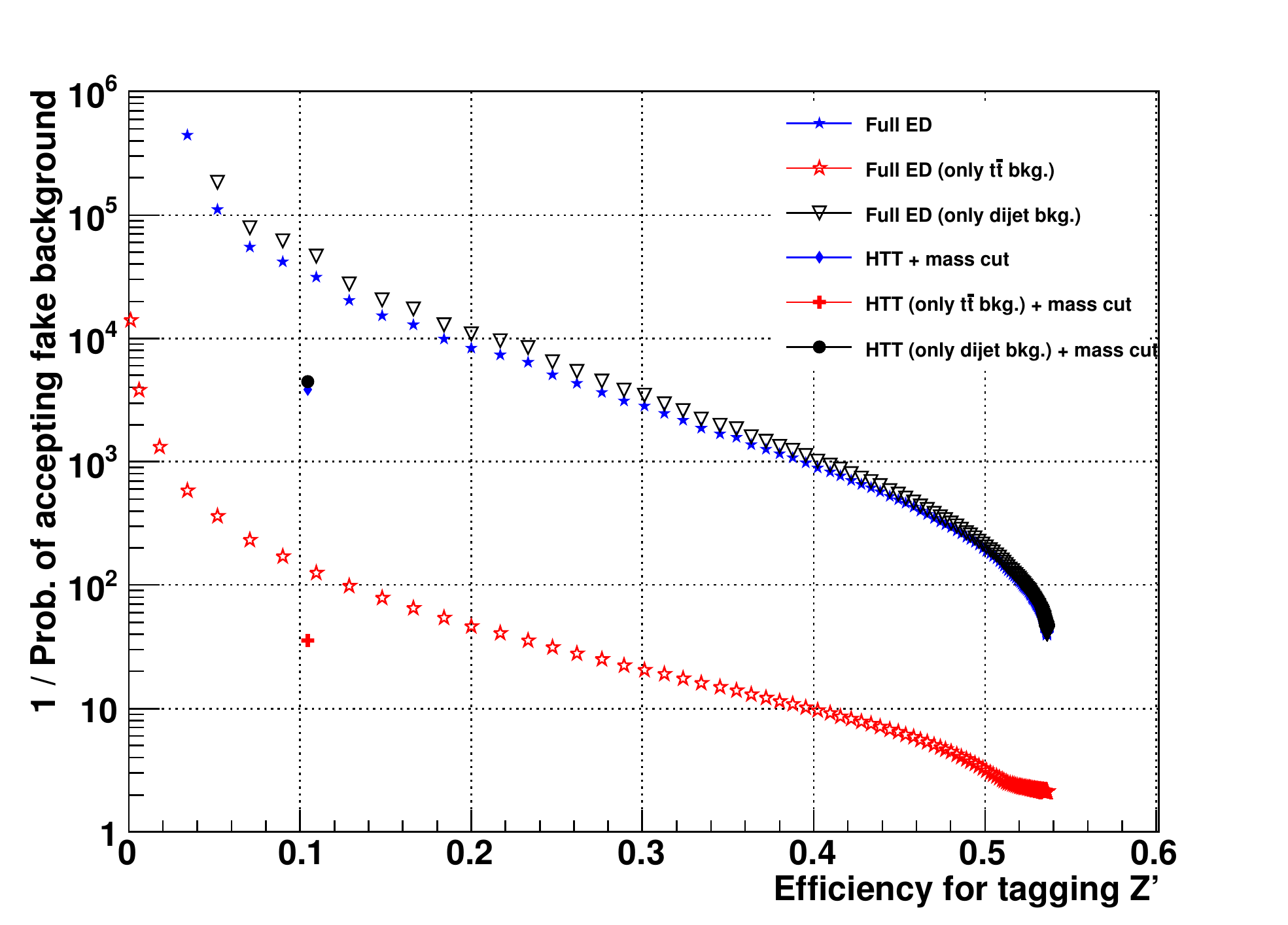}}
\caption{
Based on the $\log(\chi)$ distribution of Fig.~\ref{fig:chidists}, we show the signal acceptance versus background rejection power for event deconstruction. As a test of how much this method can improve over other methods we give performance points for the $Z'$ reconstruction based on the HEPTopTagger.  
}
\label{fig:roc}
\end{figure}
%-------------END FIGURE----------------

The variable $\chi$ can be used for a simple cut based analysis that seeks to distinguish the signal from the two backgrounds. We select events with $\chi > \chi_0$. Depending on what value of $\chi_0$ we choose, we retain a certain fraction $f(\LS)$ of the sample of signal events and a fraction $f(\LB)$ of the sample of background events (either the dijets background, the QCD $\Lt \bar \Lt$ background, or the total background). In Fig.~\ref{fig:roc}, we show the background rejection power $1/f(\LB)$ versus the signal efficiency. This plot includes the effect of events with $\chi = 0$. The black triangles show the background rejection rate for the dijets background only, the red open stars show the rejection rate for the QCD $\Lt \bar \Lt$ background, and the blue closed stars show the background rejection rate for the for the total background. The results for the total background follow closely those for the dijets background because the QCD $\Lt \bar \Lt$ cross section is much smaller than the dijets cross section\footnote{We do not use b-tagging on the fat jets or subjets. Requiring a b-tag in each fat jet would result in the two backgrounds being of similar size.}. The endpoint of the curves at $55\%$ signal efficiency is due to the $W$ and top mass window constraints. Removing those constraints would result in a curve which extends to $100\%$ signal efficiency. 

In Fig.~\ref{fig:roc}, we also compare event deconstruction with a $Z'$ reconstruction using the HEPTopTagger (HTT) \cite{HEPtagger1, HEPtagger2}. The HTT is designed to tag and reconstruct the four-momentum of boosted top quarks but is not aiming to reconstruct the full $Z'$ resonance. We consider the $Z'$ to be reconstructed if the invariant mass of the two HTT-tagged top-quark four-momenta is in the window $[1200,1600]\GeV$. We compare with the HTT because it has been used in a similar way in experimental searches in the process we study \cite{Aad:2012raa, Aad:2013gja, AtlasTop}. We find that requiring two tagged top quarks with HTT and satisfying the $Z'$ mass window cut for $m_{\bar{t}t}$ has a signal efficiency of about 0.1. Its background rejection power for the total background is about a factor 8.5 worse than that obtained with event deconstruction with this same signal efficiency. 

The difference in the performance between the two approaches could be reduced by calculating $\chi$ in Eq.~(\ref{eq:chi}) using the HTT's reconstructed top quark four momenta as input to the signal and background hard matrix elements. We refrain from doing that as this was not done in Ref.~\cite{Aad:2012raa}.

%---------------FIGURE------------------
\begin{figure}
\centerline{\includegraphics[width=11.0cm]{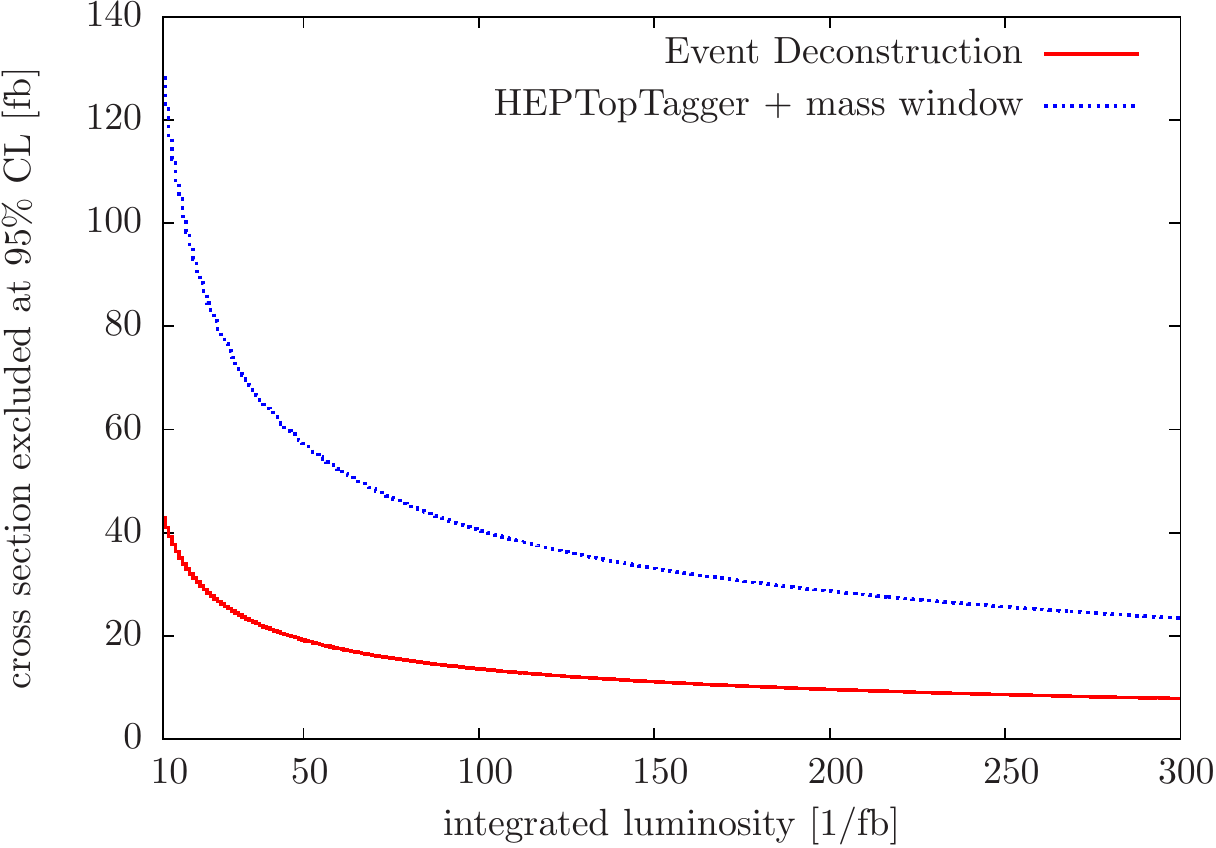}}
\caption{Signal cross section that can be excluded at the 95\% confidence level if, in fact, there is no signal found. We show the exclusion limit as a function of the integrated luminosity. The upper curve makes use of the HEP top tagger plus a mass window for the $\LZ'$. The lower curve uses event deconstruction.}
\label{fig:cs}
\end{figure}
%-------------END FIGURE----------------

We can now use the results shown in Fig.~\ref{fig:roc} to estimate what signal cross section can be excluded for a given integrated luminosity. For a given integrated luminosity $\mathcal{L}$, and a given choice of cut $\chi_0$, the number $N(\LS)$ of signal events accepted is related to the number $N(\LB)$ of background events accepted by
\begin{equation}
\label{eq:statisticalpower1}
\frac{N(\LS)}{\sqrt{N(\LB)}}
= \sqrt{\mathcal{L}}\,
\frac{\sigma(\LS) f(\LS)}{\sqrt{\sigma(\LB) f(\LB)}}
\;,
\end{equation}
where $\sigma(\LS)$ and $\sigma(\LB)$ are the signal and background cross sections, respectively. If counting statistics is the only consideration, we can rule out a signal cross section at approximately the 95\% confidence level when $N(\LS) = 2 \sqrt{N(\LB)}$:
\begin{equation}
\label{eq:statisticalpower2}
\sigma(\LS,{\rm excluded})
= 
\frac{2\sqrt{\sigma(\LB) f(\LB)}}{\sqrt{\mathcal{L}}\, f(\LS)}
\;.
\end{equation}
We can choose the signal acceptance to be $f(\LS) = 0.11$. Then Fig.~\ref{fig:roc} gives $f(\LB)$. This gives the graph of excluded signal cross section versus integrated luminosity shown in Fig.~\ref{fig:cs}. The analogous result is shown for HTT tagging of the $\LZ'$. We find that we can exclude a 3 times smaller cross section or the same cross section approximately 9 times faster using event deconstruction.

\section{Outlook and Discussion}
\label{sec:outlook}

If there is new physics to be found at the LHC, it seems likely that the new physics involves some very heavy particles that then decay into electroweak scale resonances like Z and W bosons, top quarks, or the Higgs boson, all of which have large branching ratios into hadronic jets. The transverse momentum of each of the electroweak scale resonances will then typically be larger than its mass, so that its decay products will be restricted to a limited angular region in the detector. 

It is important to look for this kind of signal, but we face the problem of separating signal events from a large background of ordinary QCD jet events. In this paper, we have proposed a method, event deconstruction, for doing this. We use a straightforward calculation of the matrix element for the hard core of the process, where the new very heavy particles appear. The hard matrix element is combined with a parton shower approximation for the decay of the electroweak scale resonances, including QCD radiation from the strongly interacting particles involved. For the background events, we again use a straightforward calculation of the matrix element for the hard core of the background process together with a parton shower approximation for the further splittings of the high transverse momentum quarks and gluons.

The idea is simple. One calculates (of course, approximately) the ratio $\chi$ of the likelihood that a given event was produced by the sought signal process to the likelihood that the event was produced by the background process or processes. This calculation is direct, based on what we know about the hard process, resonance decays, and QCD splittings. One can then test whether there is a signal by asking whether there are more events with high $\chi$ than would be likely by accident in the absence of a signal.

For both signal and background, the QCD radiation from the high $p_\LT$ partons is not a problem to be somehow minimized. Rather it provides an important clue that can help distinguish the structure of signal events from that of background events. In the analysis, the parton shower approximation is important. It includes Sudakov factors that are functions of the angle $\theta$ between two partons and that strongly modify the $1/\theta^2$ singularities of tree level perturbation theory. Furthermore, it incorporates quantum interference effects that control gluon emissions according to the color structure of the emitting partonic color antenna.

For both signal and background, the QCD radiation from the incoming beam partons {\em is} a problem to be somehow minimized. Event deconstruction does this by allowing for the possibility that some of the radiation in the angular region of interest comes from initial state radiation, which is modeled in an approximation to what a parton shower event generator does.

We have tested event deconstruction by looking for a specific signal, $\LZ' \to \Lt + \bar \Lt$. We compared the power of event deconstruction to the power of looking for two jets whose total momentum squared lies in a window around $M(\LZ')^2$, where one tags the high $p_\LT$ jets as top jets using the HEP top tagger (which is quite sophisticated and is one of the state-of-the-art methods for looking for top quark jets \cite{Aad:2012raa, Aad:2013gja, AtlasTop}). We found using event samples generated with \textsc{Pythia} that event deconstruction is more powerful than the approach based on the the HEP top tagger. This is perhaps not surprising since in earlier studies we found that the shower part of event deconstruction, by itself, is relatively efficient at identifying jets containing Higgs bosons \cite{SSHiggs} and top quarks \cite{SStop}.

The event deconstruction method is quite general and could be applied to the search for many possible physics signals. The algorithm to be applied for any one signal is rather complicated, so it might seem that the method is difficult. However, the general algorithm can be modular. For instance, our implementation, which we call \textsc{EvDec}, is constructed using C++ classes {\tt topjet}, {\tt Wjet}, {\tt gluonjet}, {\it etc}.  Thus the task of applying event deconstruction to a new process involves in large part applying an existing base of modules to the new problem.

Improvements in the methods of event deconstruction are desirable. We list a few possibilities.

\begin{itemize}

\item The basic building block of event deconstruction is the microjet: a jet defined with a very small jet radius. The simple version of event deconstruction described in this paper makes use also of a small number of `fat jets' defined using a much larger jet radius. The fat jets are then decomposed into microjets. This sort of analysis is useful when the ratio of the $p_\LT/m$ for an electroweak scale resonance is large. However, if $p_\LT/m$ is not large enough, the fat jet may not contain all of the decay products of the resonance. Additionally, in some situations one may need fat jets that overlap. In these cases, it may be better to simply eliminate the fat jets and work directly with microjets. Then one would analyze the configuration of all of the microjets in the detector, or in the central angular region of the detector. With such an analysis, one could gain accuracy in calculating $\chi$ at the expense of the speed of the calculation.

\item In our current implementation of event deconstruction, the treatment of initial state radiation is very much simplified. The initial state radiation in each fat jet is treated as being completely independent of the initial state radiation that enters other fat jets. Furthermore, the radiation from each of the two incoming partons is not distinguished from the radiation from the other. Nor is there any dependence on whether the incoming parton is a quark or a gluon. Clearly, one could gain accuracy in the description of initial state radiation at the cost of speed of the calculation.

\item At present, color connections among the partons that are the decay and splitting products of a high $p_\LT$ parton are accounted for (in the leading color approximation). These color connections influence the pattern of gluon radiation from these partons. However, the algorithms do not track the color connections between a) partons arising from one high $p_\LT$ parton and b) partons arising from a different high $p_\LT$ parton or c) partons that arose from initial state radiation or d) the two beam partons that remain after initial state radiation. One could gain accuracy by accounting for more of the color flow in each event.

\item When a high $p_\LT$ parton decays, the angles constructed from the momenta of the decay products are correlated with the spin of the decaying parton. At present, this information is used for the spin of a W-boson that was produced from a top quark decay, but it is otherwise not used. Taking better account of spin would improve the accuracy of the calculation.

\item We use two distinct calculation styles: tree level matrix elements for high $p_\LT$ partons produced at the hard interaction with large angular separations and a parton shower approximation for more soft or collinear splittings and decays. There is no sharp dividing line between where one approach is better and where the other is better. When one generates events, as opposed to analyzes events, one often uses a matching scheme to get the best of both styles of calculation at once \cite{Catani:2001cc,Alwall:2007fs}. The same sort of matching can apply to event deconstruction.

\item Event deconstruction is designed to use as much information about an event as possible. It is not immediately well adapted to situations in which invisible particles like neutrinos or photinos participate. It would be useful to extend the method to deal with invisible particles.

\item For many of the items in the list above, there is a clear path to improving the accuracy of the calculation of $\chi$. However, following this path would be costly in computational time. The basic problem is combinatorial: there are very many possible event histories that could produce the microjet configuration for a single event. In principle, one should sum over all event histories. However, if one were more clever in culling the herd of event histories, one could provide more accuracy for each one.

\item The perturbative part of the event evolution is calculated using the hard matrix element and a perturbative parton shower. There is also non-perturbative physics that is represented by a parametrized contribution to soft initial state radiation and by artificially broadened resonance widths. The soft initial state radiation can include radiation from the incoming partons as well as from the underlying event and from pileup events. Additionally, our implementation of event deconstruction includes a jet radius parameter $R$ used to define the microjets and a lower cutoff on the transverse momentum of microjets that are considered. These parameters also amount to nonperturbative inputs to the calculation. We encourage measurements by both multi-purpose experiments with the aim of optimally adjusting the nonperturbative parameters used.

\end{itemize}

From the list above, we conclude that there is a lot of room for improvement on the implementation of event deconstruction described in this paper and represented in our code, \textsc{EvDec}. We have also argued that the existing code can be extended to cover a substantial range of important problems. We plan to provide a publicly available package that includes a range of applications and, additionally, includes some of the potential improvements outlined above.

\acknowledgments{ 
This work was supported by the United States Department of Energy. We thank Danilo Enoque Ferreira, Simone Marzani and Olivier Mattelaer for valuable discussions. 

%-------------------------------------------------------------------
\end{document}